\documentclass[twocolumn,showpacs,amsmath,amssymb,aps,prl,floats]{revtex4} 

\usepackage{graphicx}
\usepackage{hyperref}
\usepackage{dcolumn}
\usepackage{bm}
\usepackage{epstopdf}
\usepackage{mathrsfs}
\usepackage{graphicx}
\usepackage{multirow}
\usepackage{amsmath}
\usepackage{threeparttable}
\usepackage{float}
\begin{document}

\preprint{arXiv:0000.00000}

\title{Mirages and Large TeV Halo-Pulsar Offsets from Cosmic Ray Propagation} 

\author{Yiwei Bao$^{1,2,3,4}$}
\author{Gwenael Giacinti$^{1,2,5}$}\email{gwenael.giacinti@sjtu.edu.cn}
\author{Ruo-Yu Liu$^{3,4}$}\email{ryliu@nju.edu.cn}
\author{Hai-Ming Zhang$^{6}$}
\author{Yang Chen$^{3,4}$}\email{ygchen@nju.edu.cn}
\affiliation{
$^1$Tsung-Dao Lee Institute, Shanghai Jiao Tong University, Shanghai 201210, China\\
$^2$School of Physics and Astronomy, Shanghai Jiao Tong University, Shanghai 200240, China\\
$^3$School of Astronomy and Space Science, Nanjing University, 163 Xianlin Avenue, Nanjing 210023, China\\
$^4$Key Laboratory of Modern Astronomy and Astrophysics, Nanjing University, Ministry of Education, Nanjing, China\\
$^5$Key Laboratory for Particle Physics, Astrophysics and Cosmology (Ministry of Education) \& Shanghai Key Laboratory for Particle Physics and Cosmology, 800 Dongchuan Road, Shanghai 200240, China\\
$^6$Guangxi Key Laboratory for Relativistic Astrophysics, School of Physical Science and Technology, Guangxi University, Nanning 530004, People's Republic of China}

\date{\today}

\begin{abstract}
The study of extended $\gamma$-ray sources usually assumes symmetric diffusion of cosmic rays. However, recent observations of multiple sources near single pulsars and significant offsets between TeV halo centroids and their parent pulsars suggest that this assumption is overly simplistic. In this Letter, we demonstrate that asymmetric propagation of cosmic rays near their accelerators may create multiple TeV sources instead of a single symmetric source. This mechanism also explains the large offsets between TeV halo centroids and their {parent} pulsars. We demonstrate that several perplexing detected sources can be naturally {interpreted} without invoking additional invisible accelerators.
\end{abstract}

\maketitle

{\it Introduction---} {The origin of cosmic rays (CRs) has been a long-standing open question for decades. Studying the extended $\gamma$-ray emission helps to reveal the identity of cosmic ray accelerators, and thus helps to uncover the origin of CRs. The traditional approach to identifying accelerators involves fitting a Gaussian to the $\gamma$-ray counts map, with the Gaussian centroid marking the accelerator's location. This method is effective in the Galactic disk, where most sources are located, and the turbulence coherence length, over which the direction of the turbulent magnetic field remains nearly unchanged, is as small as 2 pc. However, for sources located in the Galactic halo, where the coherence length is larger, it may lead to wrong predictions.} 


In reality, CR propagation is expected to be very filamentary {\cite{2012PhRvL.108z1101G,2013PhRvD..88b3010G}} on scales smaller than that of the turbulence, because these particles follow their local magnetic field lines. In contrast, standard diffusion models approximate their propagation as isotropic or anisotropic diffusion, with a homogeneous diffusion coefficient or tensor around their sources. This approximation is valid only if the diffusion length scale, the scale over which the accelerated electron loses its energy, is much larger than the coherence length $L_{\rm c}$ of the magnetic field. {The specific influence of given magnetic field structures is averaged out due} to the particle traversing many different field structures whose directions are random. As a result, in such models, the extended gamma-ray source will always be centered at the particle accelerator. In practice, this phenomenological approximation is often sufficient, because the coherence length in spiral arms {(}where most pulsars are located{)} may be as low as 2 pc \cite{2006ApJ...637L..33H}, and the lifetime of low-energy electrons is sufficiently long to allow them to cross several coherence lengths. {Therefore, for TeV halos around the Geminga and Monogem pulsars, which are located at distances of $\sim 250$ pc (and thus lie within the spiral arm) and are of $\sim 50$ pc in size, the assumption of isotropic diffusion holds, and their propagation pattern is nearly spherically symmetric. Even with large coherence length, it is still possible that the source looks symmetric due to projection effect \cite{2019PhRvL.123v1103L}.} However, challenges have now started to arise as more sources have been discovered above {$10^{1.4}$--$10^{1.6}$}\,TeV in the inter-arm regions (where the coherence length may be much larger \cite{2006ApJ...637L..33H}) and off-plane regions (where the coherence length could reach 200 pc \cite{2016JCAP...05..056B}). For such sources, the diffusion approximation is expected to break down because their electrons can only propagate within a coherence length. Hence, it is necessary to treat electron propagation in a more realistic way to explain their extended gamma-ray emissions. As shown below, if the magnetic field is bent away from or towards the observer, the electrons accumulate along the line of sight, forming apparent, off-centered gamma-ray sources, whose {orientations} on the sky may be far from that of the true parent electron source. Due to their misleading nature, we call them \lq\lq mirage sources\rq\rq\/, although the underlying physics is different from that of optical mirages. For this new type of gamma-ray sources, no astrophysical accelerators lie at their centroids, which is dramatically at odds with any naive or standard expectation. The electrons responsible for the detected sources have, in fact, been accelerated in a nearby accelerator, which may even be located outside the extent of the detected gamma-ray source.

{HAWC \cite{2020ApJ...905...76A} and LHAASO \cite{2024ApJS..271...25C} have detected almost a hundred $\gamma$-ray sources at TeV energies and beyond. Many of these sources are extended and likely due to inverse Compton (IC) from relativistic electrons. One prominent and intriguing example of such leptonic sources is the ``TeV halo''. TeV halos are extended gamma-ray emitting regions around middle-aged pulsars. They are due to electrons with energies up to about sub-PeV\cite{2014ApJ...783L..21S,2020A&A...642A.204C,2021Sci...373..425L,2022IJMPA..3730011L,2022A&A...666A...7M}, and with typical spatial extensions of order $\approx50$ pc. Within these regions, the diffusion coefficient of CRs is suppressed by a factor of $10^2$--$10^3$ compared with the value derived from the Boron-to-Carbon ratio \cite{2017Sci...358..911A}. Studying TeV halos not only provides insights into these unique celestial phenomena but also helps to understand the local magnetic field \cite{2018MNRAS.479.4526L} and explain the excess in the diffuse $\gamma$-ray background \cite{2024NatAs...8..628Y}. Here, we take pulsar halos as an example to illustrate our theory on asymmetric diffusion, and the theory itself could be applied to other types of sources.}

In this Letter, instead of adopting the traditional diffusion approximation to investigate electron propagation around their sources, we calculate from first principles the motion of individual electrons in synthetic three-dimensional turbulence. The latter approach can give vastly different results, and it is more realistic because the (effective) cosmic-ray diffusion tensor strongly varies in space on length scales that are smaller than the coherence length. Our method is more fundamental and more flexible than using the diffusion approximation. Indeed, {It can resolve the filamentary propagation of electrons on small scales, while reproducing the predictions of the diffusion approximation on large scales.} By considering the short lifetime of electrons {of $\sim 100$ TeV, whose diffusion-loss scale is $\sim 70$ pc under a magnetic field of $3 \mu$G}, we find that the real distribution of these electrons around their sources can be very filamentary even in strong turbulent magnetic fields. Also, we find that both the presence of large offsets from the pulsars and the presence of mirage sources are two interlinked phenomena arising from this filamentary propagation. Our results suggest that in the inter-arm region and the Galactic halo where the coherence length is large, a large pulsar halo could be identified as multiple distinct smaller halos. Moreover, when the primary identified halo exhibits low luminosity, {it may drown in the background noise} these ancillary mirage halos could emerge as the sole identified entities, leading to considerable offsets between the detected gamma-ray sources and the actual astrophysical source of the electrons.

{\it Methods---} 
We use test particle simulations, where we integrate the trajectories of electrons {of $\sim 100$ TeV} in a turbulent magnetic field. {The turbulent magnetic field is generated with four layers of nested grids \cite{2012JCAP...07..031G,2013PhRvD..88b3010G, 2020Ap&SS.365..135M}, which are responsible for four different scales (CRProp3 and Pencil codes have the turbulence generation modules for these layers).} For the trajectory integrator, we use the original Boris pusher \cite{2018PhPl...25k2110Z}. These electrons display an IC spectrum mainly off the cosmic microwave background, peaking around {2}0 TeV. The electrons are injected at {(0,0)} in the simulation, and with a power-law spectrum ${\rm d}N/{\rm d}\gamma \propto \gamma^{-\alpha} \exp(-\gamma/\gamma_{\rm cut})$, where $N$ denotes the electron number, {$\alpha \approx 2$ the power-law index, and $\gamma_{\rm cut}=2\times 10^9$ the cutoff energy. {In practice, we inject a single numerical particle per logarithmic energy bin (8192 bins) and assign each a weight proportional to $E^{1 - \alpha} \exp(-\gamma / \gamma_{\rm cut})$, where the factor $E^{1-\alpha}$ arises from the transformation ${\rm d}N/{\rm d}\log E = E\,{\rm d}N/{\rm d}E$. We inject all
 electrons at $t=0$, propagating them and cooling them with synchrotron and IC losses, recording their positions every 10 years, and each recording can be considered as a new particle.} The dependence of morphology on power-law index and cutoff energy is relatively weak. The spin-down luminosity of the pulsar is approximated to be a constant \cite{2018MNRAS.479.4526L}.} We only study here sources located within a few kpc from the Sun, due to their better detectability. Therefore, for the seed photon energy density, we use the interstellar radiation field averaged over a 2\,kpc region surrounding the Sun.

For the simulation data analysis, we assume that the TeV halos generated by {such} pulsars are observed by LHAASO KM2A \cite{2021ChPhC..45b5002A}. 
We consider the observations to be at {$10^{1.4}$--$10^{1.6}$}\,TeV, and we set here the reference luminosity of the pulsar halo {as} $L_{\rm ref}=6\times 10^{32}\,$erg s$^{-1}$. This $\gamma$-ray luminosity is about 6 times that of the observed luminosity of Geminga pulsar's halo in the range of $8-40$\,TeV \cite{2017Sci...358..911A}, but still reasonable. A wider range of luminosities is explored in \cite{paperII} (paper II). For fainter sources, the lower luminosity can be compensated by an increase in observation time. Assuming all arriving photons to be at {2}0\,TeV, the resulting annual {number of} counts is $8.1\times 10^5(250\,{\rm pc}\,d^{-1})^2$ (with $d$ being the distance to the pulsar). Our background is set to be consistent with that in the Crab Nebula region, which results in 1{0} event per hour within a 1$^\circ$ cone \cite{2021Sci...373..425L}. We set the observation time to 3\,yr, which is roughly the LHAASO effective operation time, and we assume that the source is visible for 6 hours per day. {The counts map are calculated in the band {$10^{1.4}$--$10^{1.6}$} TeV.}

To mimic the influence of the instrument's point spread function (PSF), we smooth the obtained image by convolving it with a 2D symmetric Gaussian function with {$\sigma_{\rm G}=\sigma/1.51=0.3^\circ$}({corresponding to} LHAASO KM2A resolution at {$10^{1.4}$--$10^{1.6}$} TeV \cite{2021ChPhC..45b5002A, 2024ApJS..271...25C}). The PSF-convolved counts map is further binned into grids of $0.1^\circ \times 0.1^\circ$, as per LHAASO's analysis. We employ the standard \lq\lq on-off method\rq\rq\/ to analyze the mock data. 

{Our simulations use synthetic turbulent magnetic fields based on Ref.\cite{1999ApJ...520..204G}, and assume isotropic turbulence with a Kolmogorov power spectrum. However, the properties of the interstellar turbulence may be different, see e.g. \cite{2002PhRvL..89B1102Y,2024ApJ...965...65M}. 

Although realistic MHD turbulence is intermittent, the intermittency may enhance the mirage effect because the intermittent turbulence has coherent structures and
local spikes (see e.g., Ref \cite{2023PhRvD.108j3029C}, Fig. 1 for a comparison between real turbulence and the Gaussian turbulence we use).

The simulation itself has very good resolution and is accurate at the particle gyration radius. The mean free path of 1 PeV particles in 3 $\mu$G magnetic field is $\sim 1$ pc, much smaller than the source separation, so the particles are well scattered \cite{2018JCAP...07..051G}.
}

{\it Multiple sub-halos and large TeV halo-pulsar offsets---} 
In our simulations, we frequently ``detect'' multiple halo-like sources forming near a single accelerator. The underlying reason for having such multiple sources is that when the electrons diffuse along magnetic field lines, the lines bend on scales comparable with the coherence length. When these bent field lines align roughly with our line of sight, a projection effect occurs, causing the electrons to accumulate along the line of sight and to produce a mirage source.

The offsets observed in the data between TeV halo centroids and pulsars locations can also be attributed to this phenomenon of mirage sources. More precisely, if the primary halo that is closer to the pulsar happens to be dimmer than its adjacent {ancillary} halos, the main halo may become elusive, and might even drown in the background noise. As a result, the only ``detectable'' halo could be the brighter adjacent one, resulting in a surprizingly large apparent offset between the detected TeV halo and the location of its parent pulsar.

In paper II, we investigate how properties of a mirage source are influenced by the regular-to-turbulent magnetic field strength ratio. Provided that {the $B_{\rm r}/B_{\rm t}$ ratio} is $\lesssim 1$ (we explore up to 3, since a much higher ratio would not be realistic in the ISM), it does not significantly impact the properties of mirage sources. Therefore, in the following, we will not further discuss the impact of $B_{\rm r}/B_{\rm t}$.

\begin{figure}
    \centering
\includegraphics[width=0.44\textwidth]{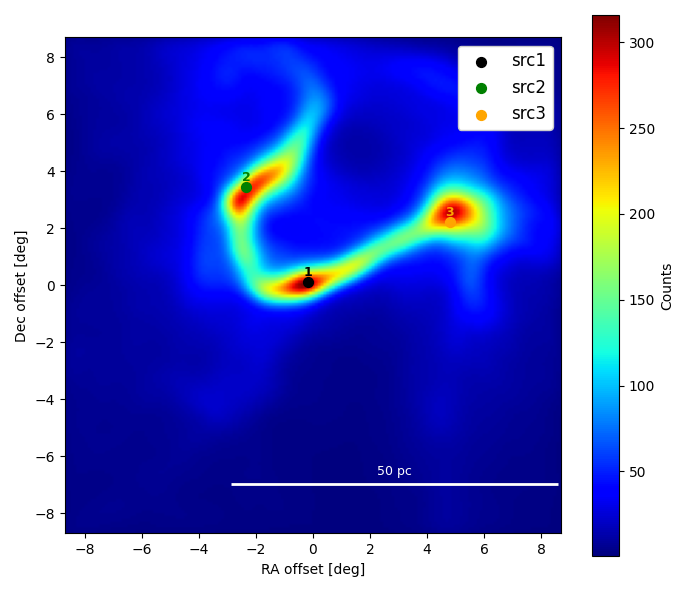}\\
\includegraphics[width=0.44\textwidth]{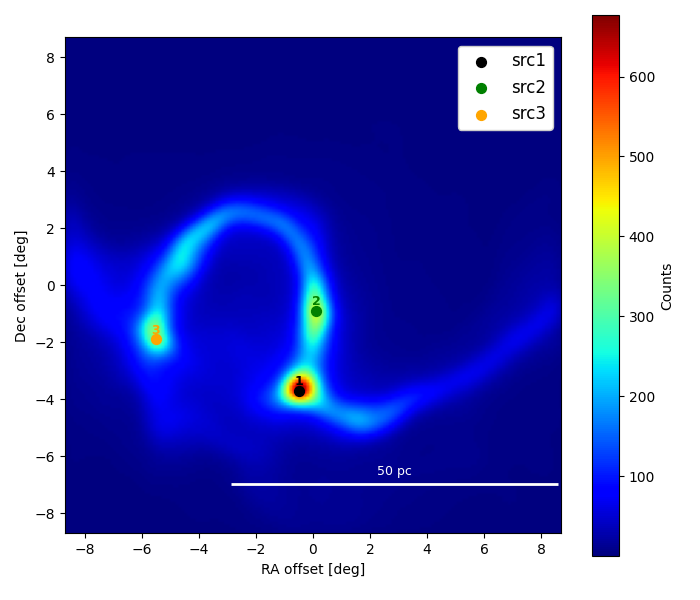}
    \caption{A single halo could be misinterpreted as triple halos. {The distance to the pulsar is 250 pc, $L_{\rm c}=40$\,pc, $B_{\rm r}=0$ and $B_{\rm t}= 3\,\mu$G.}
    The color indicates the number of counts in each grid point (smoothed with Gaussian PSF). The noise has been removed. The particles are injected at (0,0).}\label{fig:3source}
\end{figure}

In \autoref{fig:3source}, the particles are injected at the center of each panel at (0,0), in two different realisations of the magnetic turbulence with $L_{\rm c}=40$\,pc, $B_{\rm r}=0$ and $B_{\rm t}= 3\,\mu$G. {the colored points represent the best Gaussian fits of the three sources.} The distance of the pulsar is set to 250\,pc. The colors represent the gamma-ray emission from these sources, and the noise is not displayed. One can see that a source forms near the injection site in each panel (orange dot for its centroid). Along the filamentary magnetic field, we also ``detect'' two other sources with similar brightness (black and green dots). Therefore, a single large halo can be very filamentary and can potentially be identified as triple Gaussian sources lying along the filamentary structure.
 

\begin{figure}
    \centering
{\includegraphics[width=0.4\textwidth]{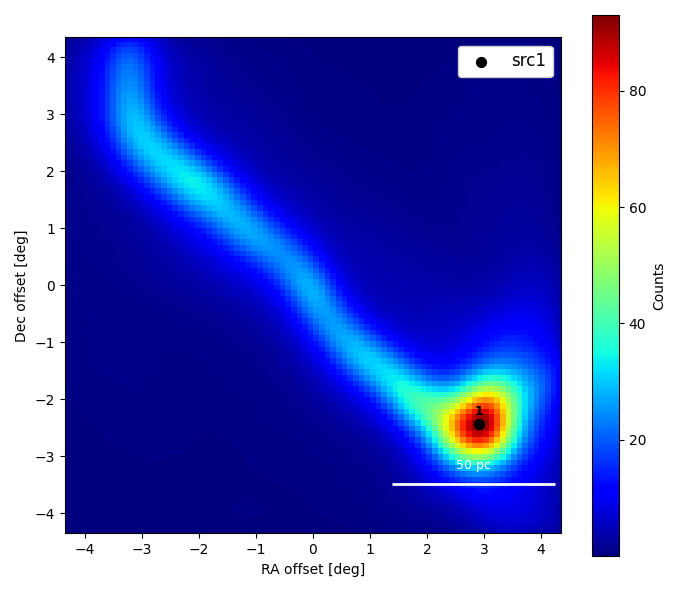}}\\
{\includegraphics[width=0.4\textwidth]{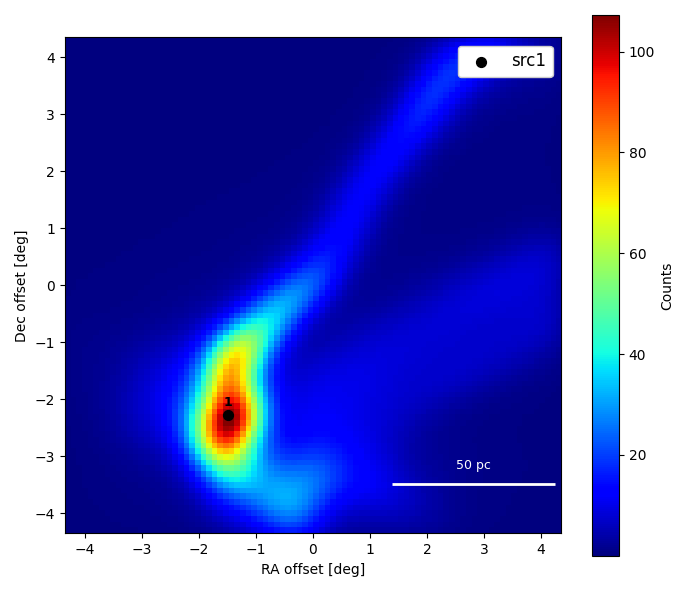}}
    \caption{Cases in which large offsets can arise from the same mechanism. The color representation indicates the number of counts in each grid point. The noise has been removed. Parameters are set to $L_{\rm c} = 200$\,pc (mean Galactic value~\cite{2016JCAP...05..056B}), $B_{\rm r}=0$ and $B_{\rm t}= 0.5\,\mu$G (reasonable, see table of paper II).}\label{fig:offsets1}
\end{figure}

In \autoref{fig:offsets1}, two configurations with very large offsets are displayed. As can be seen, the main sources near the parent pulsars could have half of the surface brightness of the secondary sources. With noises taken into consideration, we may only see the {adjacent ancillary sources} sources, with the main sources being drowned in the noise. In these two panels, we put the sources at a distance of $1\,$kpc. {The offsets could be still significant at $\sim 3$ kpc.} {Also, as can be seen from \cite{paperII}, the mirage ratio is negatively correlated with $d$, $B_{\rm r}/B_{\rm t}$ and $B^2$, and positively correlated with $L_{\rm c}$ (strong correlation) and $\alpha$ (weak correlation).}

\begin{figure}
    \centering
\includegraphics[width=0.44\textwidth]{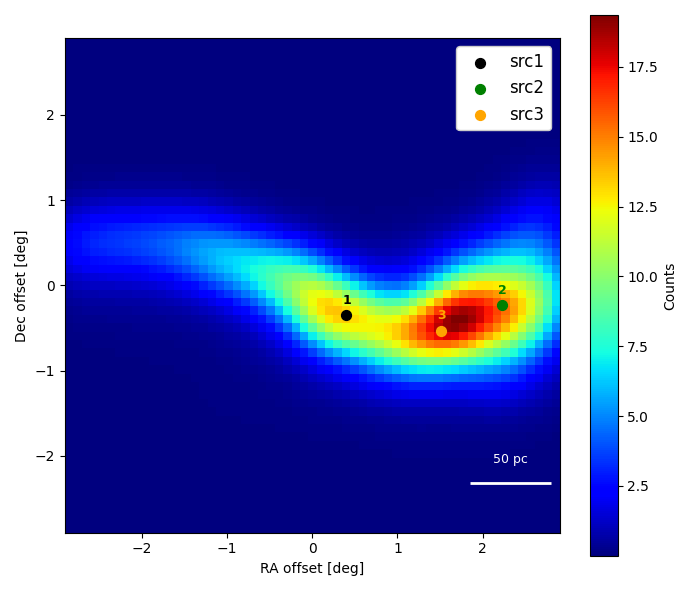}\\
\includegraphics[width=0.44\textwidth]{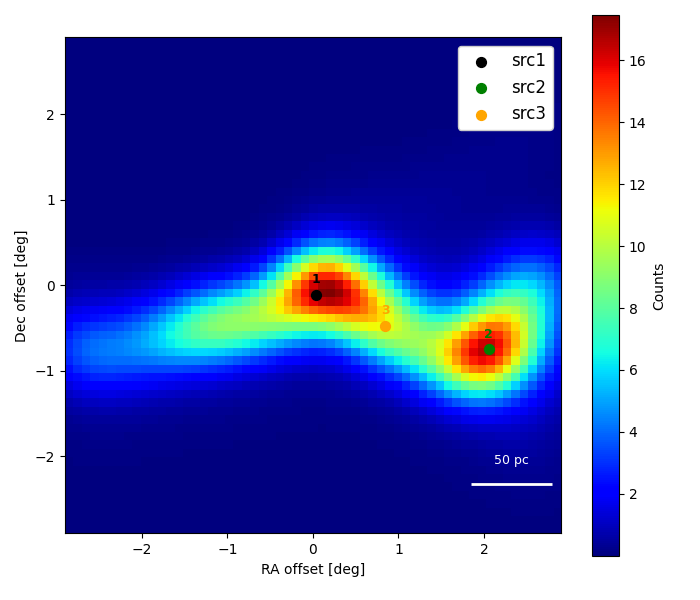}
    \caption{Two configuration which can {produce analogous shapes of} the 1LHAASO J0206+4302u and 1LHAASO J0212+4254u, with the color represent in the number of counts in each grid point. The noise has been removed. {$L_{\rm c} = 200$\,pc, and all injected} electrons have an initial positive {x} velocity to mimic the particle injection at the tail of BSPWN, and the phenomenon is more obvious with narrower injection. Parameters are set to $B_{\rm r}=0.7\,\mu$G and $B_{\rm t}=0.7\,\mu$G, and the distance is set to be $d=3$ kpc (distance of the milli-second pulsar).}
\label{fig:333}
\end{figure}

{\it Applications---} Despite their complicated evolution in their early ages, a large fraction of pulsar wind nebulae (PWNe) will ultimately evolve into a bow shock pulsar wind nebula (BSPWNe) due to their {supersonic} kick velocities ($100$--$500$\,km s$^{-1}$ \cite{2017A&A...608A..57V}) {in the interstellar medium (ISM), especially after they travel outside their parent supernova remnants}. 
In \autoref{fig:333}, we show {two} configurations which can reproduce the sources 1LHAASO J0206+4302u and 1LHAASO J0212+4254u. The offset from 1LHAASO J0206+4302u to the millisecond pulsar J0218+4232 can be attributed to the BSPWN whose tail can extend for tens of pc before it gets destroyed by neutral mass loading \cite{2012ApJ...746..105N,2015MNRAS.454.3886M}. The real separation of the two sources is $\approx 70$\,pc. In {all panels}, $L_{\rm c} = 200$\,pc, and all electrons have an initial positive {x} velocity to mimic the particle injection at the tail of the BSPWN. The phenomenon is more obvious with narrower injection. The total $\gamma$-ray luminosity is roughly 10 times that of Geminga halo. Parameters are set to $B_{\rm r}=0.7\,\mu$G and $B_{\rm t}=0.7\,\mu$G (see paper II for the turbulent and regular field strength at the location of PSR J0218+4232). The distance is set to be $d=3$ kpc ({i.e.,} distance of the milli-second pulsar). Meanwhile, Faraday rotation measurements have revealed a large-scale magnetic field parallel to the Galactic plane \cite{2024ApJ...966..240X}, which is consistent with our study here because 1LHAASO J0206+4302u and 1LHAASO J0212+4254u lie almost parallel to the Galactic disk. 

Using this model, the offsets observed in many TeV halos can be naturally accounted for, such as J2005+3415 and J2005+3050 (see paper II for detailed calculations).


{\it Discussion---} 
Even though mirage sources look similar to real sources, it is possible to identify them as mirages. The most intrinsic feature of a mirage source is that there is always a connecting structure in gamma-rays (probably slightly dimmer than the sources in TeV) {between} the pulsar and the mirage source. Although with current observations the connecting structure may be drowned in the noise, it might be possible to detect it in the future with a larger observation time, or with experiments with better sensitivities.

It is also possible to distinguish mirage halos thanks to observations in other wavelengths. The magnetic flux tube in the halo is oriented nearly along the line of sight, resulting in the synchrotron emission from the electrons {that is} mainly emitted perpendicular to the line of sight. Therefore, mirage sources are expected to be very faint in X-rays, compared to their $\gamma$-ray luminosity \cite{2019PhRvL.123v1103L}. On the other hand, the thin structure connecting the primary source to the mirage sources should be brighter in X-rays and easier to detect, also because of the orientation of the magnetic fields. In contrast, if the gamma-ray sources are {powered by invisible in situ radio-quiet pulsars}, there should be no structures connecting the pulsars and the offseted sources, and there will be a small X-ray PWN surrounding the very vicinity of the pulsar, just like the Geminga PWN \cite{2017ApJ...835...66P}. Therefore, future X-ray telescopes with low sensitivities could help confirm whether the source is a mirage, or is contributed by a radio-quiet invisible pulsar.

When the source is bright, the asymmetry increases the number of detected sources, but when the source is dim, the asymmetry decreases the number of detected sources because it is harder to identify a source with a standard Gaussian if the source is faint and far away ($\sim 3$ kpc) from us.

\cite{2012PhRvL.108z1101G,2013PhRvD..88b3010G}, as well as the formation of mirage sources and large offsets. Our method can handle a wide range of turbulence length scales within reasonable computing times. More realistic simulations, {such as} particle-in-cell or 3D magnetohydrodynamic simulations, only cover limited spatial scales, restricting the range of electron energies that can be studied with them.

{\it Conclusions and perspectives---} 
The assumption of isotropic cosmic-ray diffusion, which is traditionally used in the study of TeV halos around pulsars, is getting challenged by recent observations of multiple sources near single pulsars and significant offsets between TeV halo centroids and their parent pulsars. In this Letter, we introduced the novel concept of mirage sources. We demonstrated, by propagating individual electrons from first principles, that the intrinsically asymmetric propagation of electrons around pulsars can create multiple detectable TeV sources instead of a single symmetric TeV halo. This mechanism also explains large offsets between TeV halo centroids and their parent pulsars.

We showed that many perplexing gamma-ray sources can be interpreted through asymmetric propagation very naturally. In particular, one asymmetric halo generated by a pulsar may be identified as multiple sources. 
Our model explains observed halo-like TeV $\gamma$-ray emissions with significant offsets from nearby pulsars. It suggests that these halos may be attributed to those pulsars despite their significant offsets, rather than by invisible radio-quiet pulsars closer to the TeV halo centers. 
Our results are also consistent with the large scale magnetic field found parallel to the Galactic disk \cite{2024ApJ...966..240X}. Our model predicts that new TeV halos discovered at high Galactic latitude, if any, will likely show a filamentary structure and maybe also mirage sources/large offsets.

\begin{acknowledgments}
We thank Chunkai Yu, Hao Zhou, {Xiang-dong Li} and Giovanni Morlino for discussion. This work is supported by NSFC under grants No. 12393853, 12350610239, 12393852, 12173018, 12121003 and U2031105, {by the Fundamental Research Funds for the Central Universities under No. 020114380057, and by K. C. Wong Educational Foundation.}
\end{acknowledgments}
\bibliography{apssamp}

\end{document}